# Grouping the executables to detect malwares with high accuracy

Sanjay K. Sahay[a], Ashu Sharma[b,*]

[a]Assistant Professor, Dept of Computer Science and Information System, BITS PILANI, K. K. Birla Goa Campus, India
[b]Research Scholar, Dept of Computer Science and Information System, BITS PILANI, K. K. Birla Goa Campus, India

**Abstract**

The metamorphic malware variants with the same malicious behavior (family), can obfuscate themselves to look different from each other. This variation in structure lead to a huge signature database for traditional signature matching techniques to detect them. In order to effective and efficient detection of malwares in large amounts of executables, we need to partition these files into groups which can identify their respective families. In addition, the grouping criteria should be chosen such a way that, it can also be applied to unknown files encounter on computer for classification. This paper discusses the study of malwares and benign executables in groups to detect unknown malwares with high accuracy. We studied sizes of malwares generated by three popular second generation malwares (metamorphic malwares) creator kits viz. G2, PS-MPC and NGVCK, and observed that the size variation in any two generated malwares from same kit is not much. Hence we grouped the executables on the basis of malware sizes by using Optimal k-Means Clustering algorithm and used these obtained groups to select promising features for training (Random forest, J48, LMT, FT and NBT) classifiers to detect variants of malwares or unknown malwares. We find that detection of malwares on the basis of their respected file sizes gives accuracy up to 99.11% from the classifiers.

## 1. Introduction

As new variants of malwares getting evolve every day, malwares defense becoming increasingly difficult task in detecting malware and protecting computers systems from them[1]. Recently 11 zero-day vulnerabilities reported during the month of August while 6 of these were reported in industrial control systems[2]. Even state sponsored highly skilled hackers are developing customized malwares to disrupt industries and for military espionage[3]. Many of countries continue to incur most costly data breaches. Among them two countries had the highest cost from data breach[4] (the U.S. at $5.4 million and Germany at $4.8 million).

Anti-malware industries are facing a major challenge of continuously increase of huge data, which need to be checked for potential malicious content. Microsoft reports that there real-time detection anti-malware products are present on over 160 Million computing devices throughout the globe and they daily analyze tens of millions of data files as potential malware[5]. Reason behind these high volumes of different files is that the malware authors introduce metamorphism to the malicious components. Metamorphic malware represent the next class of virus that can create an entirely new variant after reproduction[3]. The new variant produced is in no-way similar to the original variant which lead a huge increase in the malwares count.

In order to detect them with high accuracy, we need to group them to identify their respective families. In addition, such grouping criteria may be applied to new test executables to classify it to malware. In this paper we studied three popular second generation malwares creator kits viz. G2, PS-MPC and NGVCK and found that the size variation in any two generated malwares from same kit does not differ much. Hence in this work we grouped the executables on the basis of malware sizes by using optimal k-Means Clustering algorithm and promising features are selected separately from each groups. Further these obtained features are tested on random forest, J48, Logistic Model Trees (LMT), functional trees (FT) and naive bayes tree (NBT) classifier using machine learning technique.

[*] Corresponding author. Tel.: +91-8975805861.
*E-mail address:* p2012011@goa.bits-pilani.ac.in

The paper is organized as follow, in next section related work is discussed, In section 3 we present our approach, The section 4 discuss the experimental results and finally section 5 contains the conclusion and future directions.

## 2. Related work

The first virus was created in 1970[6] and since then there is a strong contest between the attackers and defenders. To combat the threats/attacks from the second generation malwares, Schultz et al. (2001) was the first to introduce the concept of data mining to classify the malwares[7]. In 2005 Karim et al.[8] addressed the tracking of malware evolution based on opcode sequences and permutations. O. Henchiri et al.(2006) proposed a hierarchical feature extraction algorithm and used ID3, j48, Naive Bayes and SMO classifier and obtained maximum of 92.56% accuracy[9]. In year 2005, Karim et al.[8] addressed the tracking of malware evolution based on opcode sequences and permutations. O. Henchiri et al.(2006) proposed a hierarchical feature extraction algorithm and used ID3, j48, Naive Bayes and SMO classifier and obtained maximum of 92.56% accuracy[9]. Bilar (2007) uses small dataset to examine the opcode frequency distribution difference between malicious and benign code[10] and observed that some opcodes seen to be a stronger predictor of frequency variation. In 2008, Yanfang Ye et. al.[11] applied association rules on API execution sequences for classifying the malwares. In 2008, Tian et al.[12] classified the Trojan using function length frequency and shown that the function length along with its frequency is significant in identifying malware family and can be combined with other features for fast and scalable malware classification. Moskovitch et al.[13] compared the different classifiers by byte-sequence n-grams (3, 4, 5 or 6). Among the classifiers they studied BDT, DT and ANN out-performed NB, BNB and SVM classifiers, exhibiting lower false positive rates. In year 2008, Siddiqui et al.[14] used variable length instruction sequence for detecting worms in the wild. They tested their method on a data set of 2774 (1444 worms and 1330 benign files) and got 95.6% detection accuracy. In 2009 S. Momina Tabish[15] used 13 different statical features computed on 1, 2, 3 and 4 gram by analyzing byte-level file content for classification of malwares. In year 2010, Bilal Mehdi et. al.[16] used hyper grams (generalized n-gram) and obtained 87.85% detection accuracy and claimed no false alarm. Chatchai Liangboonprakong et al. (2013) proposed a classification of malware families based on N-grams sequential pattern features[17]. They used DT, ANN and SVM classifier and obtained good accuracy. Santos et al. in year 2011 pointed out that supervised learning requires a significant amount of labeled executables for both malware and benign programs, which is difficult to obtain, hence they proposed a semi-supervised learning method for detecting unknown malwares, which does not require a large amount of labeled data[18]. They obtained 86% of accuracy by labeling only 50% of the selected data set. In subsequent paper[19] in 2013, they used Term Frequency for modeling different classifiers and found that SVM outperform with accuracy of 92.92% and 95.90% respectively for one opcode and two opcode sequence length respectively. Recently (2014) Zahra Salehi et al. construct feature set by extracting API calls used in the executables for the classification of malwares[20].

## 3. Our approach

Fig. 1 represents the procedure to partition the dataset in groups, finding the promising features from each formed group and the classification of unknown malwares by using Random forest[21], J48[22], LMT[23], FT[24] and NBT[25] classifiers available in WEKA tool[26].

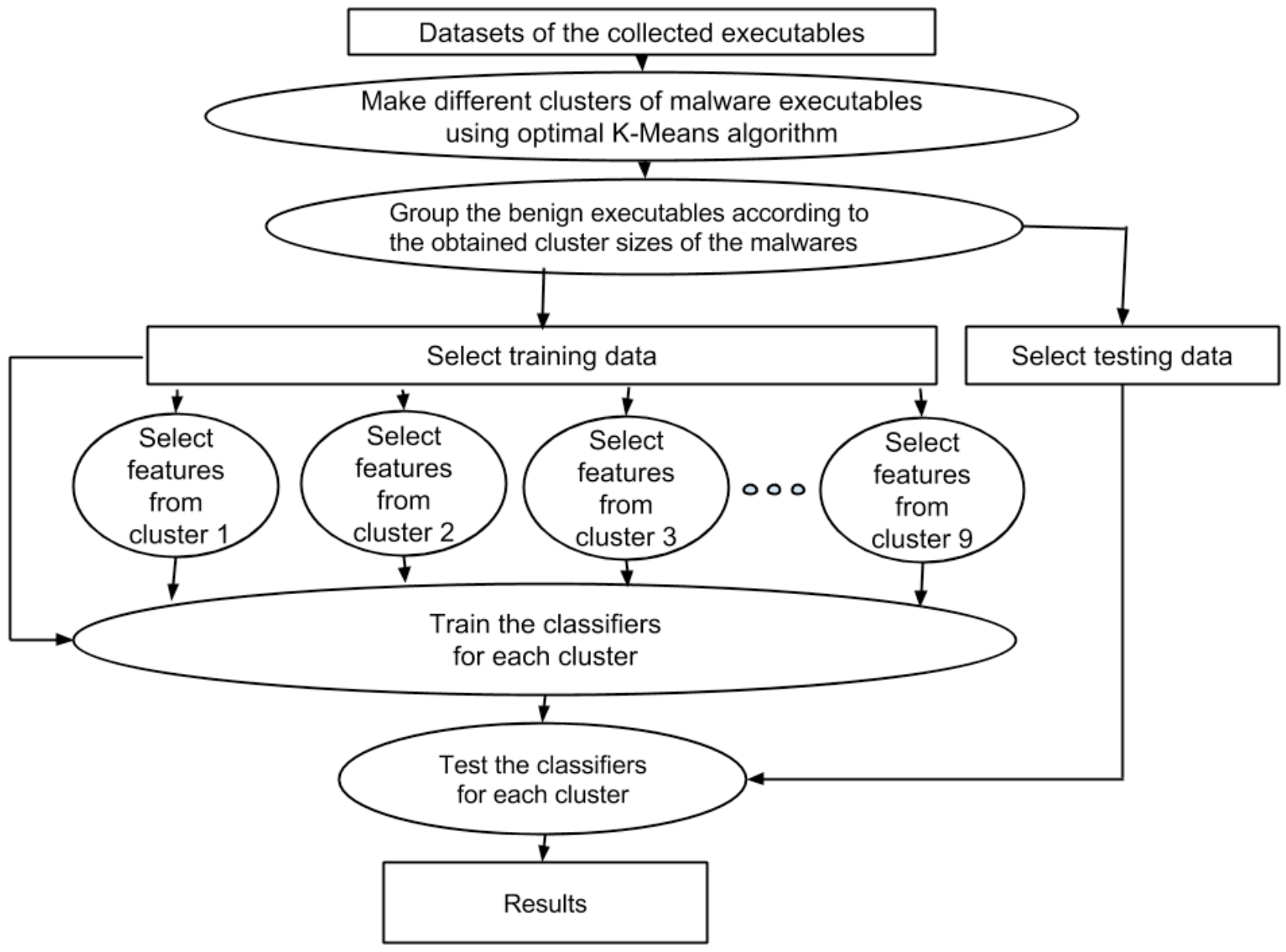

Figure 1: *Flow chart for the proposed method.*

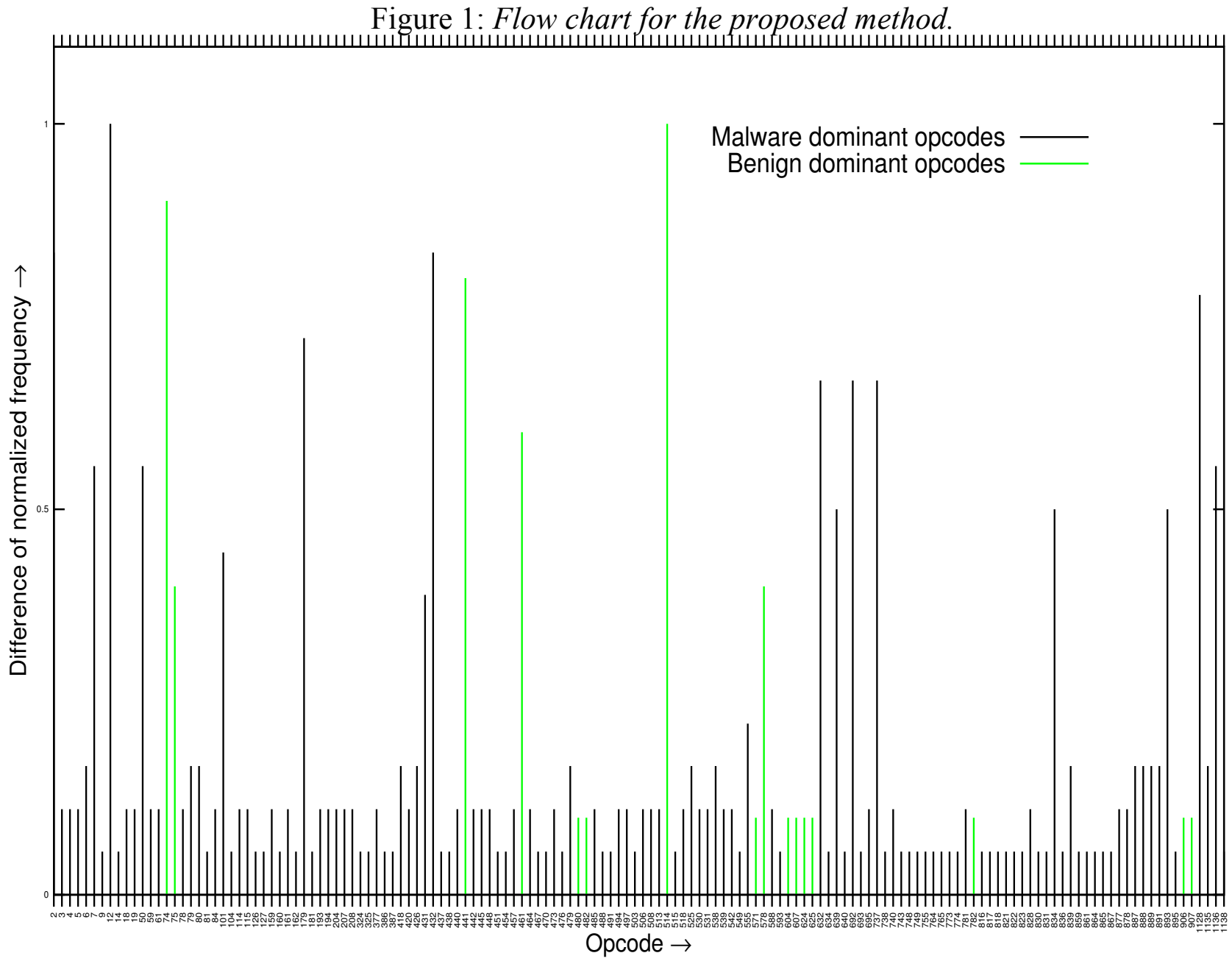

Figure 2: *Opcodes that have high difference of occurrence in malware and benign executables.*

*3.1.*

*3.2. Data preprocessing*

For the experimental analysis, we downloaded 11368 malwares from malacia-project[27] and collected 3151 benign programs (also verified from virustotal.com[28]) from different systems.

**Algorithm 1:** Selection of the Promising Features

**INPUT:** Malwares and benign assembly codes, $\mathbf{N_b}$ → Total No. of benign in the cluster, $\mathbf{N_m}$ → Total No. of malware in the cluster, **N** → Required No. of features.
**OUTPUT:** Features for the analysis.
**BEGIN**
**for all** Malwares and benign data **do**
Compute the sum of Normalized frequency($\mathbf{f_i}$) of each opcode $\mathbf{o_j}$.

$$S_{F_m}(o_j) = \left(\sum f_i(o_j) / Max(f(o_j))\right) / N_m$$

$$S_{F_b}(o_j) = \left(\sum f_i(o_j) / Max(f(o_j))\right) / N_b$$

**end for**
**for all** opcode $\mathbf{o_j}$ **do**
Compute the difference $\mathbf{D(o_j)}$ between the $S_{F_m}(o_j)$ and $S_{F_b}(o_j)$ for each opcode.

$$D(o_j) = \left|S_{F_m}(o_j) - S_{F_b}(o_j)\right|$$

**end for**
**return N** number of opcodes with highest **D**(**o**).

For the analysis we disassemble all collected executables to their assembly codes by *objdump* utility available in the Linux system and found 1147 unique opcodes, which we labeled with a fixed integer. viz. 1 ← aaa, 2 ← aad, ...., 1147 ← xsha256.

We study the obtained opcodes occurrence in malwares and benign executable and found that many opcodes occurrences in malwares differ significantly from benign program, and vice versa (fig. 2). Hence, we obtained the promising features by computing the difference of normalized opcodes frequency between benign and malware executables as given in the algo. 1.

*3.3. Partitioning the executables in dataset*

We studied the malwares sizes generated by popular metamorphic malware generator kits viz. NGVCK[29], PS-MPC[30] and G2[31] and found that size of the malwares generated by any one of the kits does not vary much (fig. 3). For efficient analysis and classification of advanced malwares in large amounts of executables, we used size of an executable as criteria for grouping the dataset. We partitioned the collected dataset into 9 groups by using optimal k-Means Clustering algorithm. The number of clusters (value of K) is obtained by the Bayesian information criterion[32].

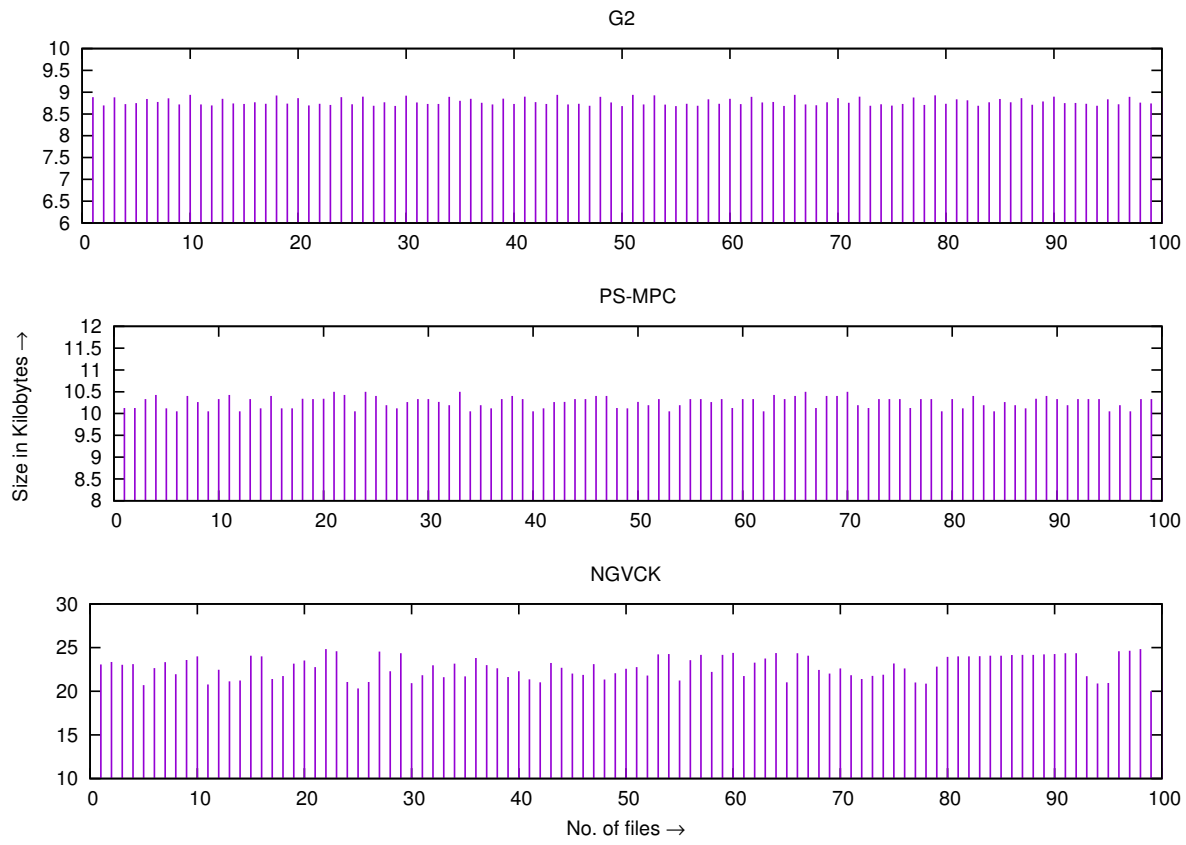

Figure 3: *Variation in the size of malwares generated by three different malware generators.*

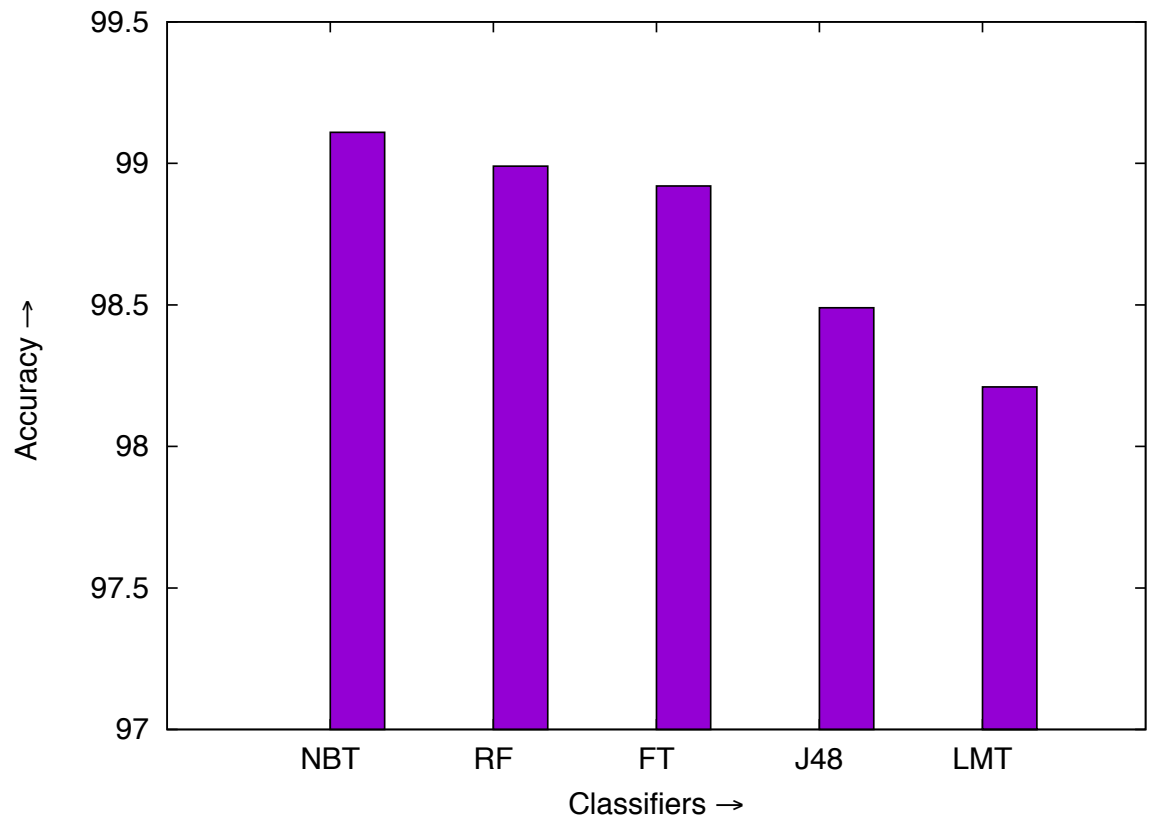

Figure 4: *Detection accuracy by selecting most promising number of features*

## 4. Experimental analysis and results

We have used machine learning to train and test the Random forest, J48, LMT, FT and NBT classifiers available in WEKA tool (a collection of visualization tools and algorithms for data analysis and predictive modeling, together with graphical user interfaces for easy access to these functionality[26]).

Table 1. Number of executables in each cluster of malwares and benign dataset.

| Cluster | No. of malwares for training | No. of benign for training | No. of malwares for testing | No. of benign for testing |
|---|---|---|---|---|
| 1 | 322 | 43 | 55 | 18 |
| 2 | 1234 | 20 | 221 | 9 |
| 3 | 1489 | 20 | 265 | 9 |
| 4 | 714 | 71 | 128 | 13 |
| 5 | 335 | 2227 | 61 | 402 |
| 6 | 886 | 36 | 158 | 11 |
| 7 | 2716 | 40 | 481 | 11 |
| 8 | 1148 | 33 | 204 | 11 |
| 9 | 18 | 156 | 4 | 21 |
| Total | 8862 | 2646 | 1577 | 505 |

We have obtained 9 groups from the datasets, which are further divided into two sets. One of the sets is used to training the classifiers and other one is used to test the detection accuracy of trained classifiers. The training set consists of 8862 malwares and 2646 benign executables and for testing the classifiers 1577 malware 505 benign executables are taken. The number of malwares and benign executables for training and testing the classifiers are given in table 1. In this for robust results, we ensure that at least 15% of the executables in the cluster that are not used for training purpose, are taken for the testing of the classifiers.

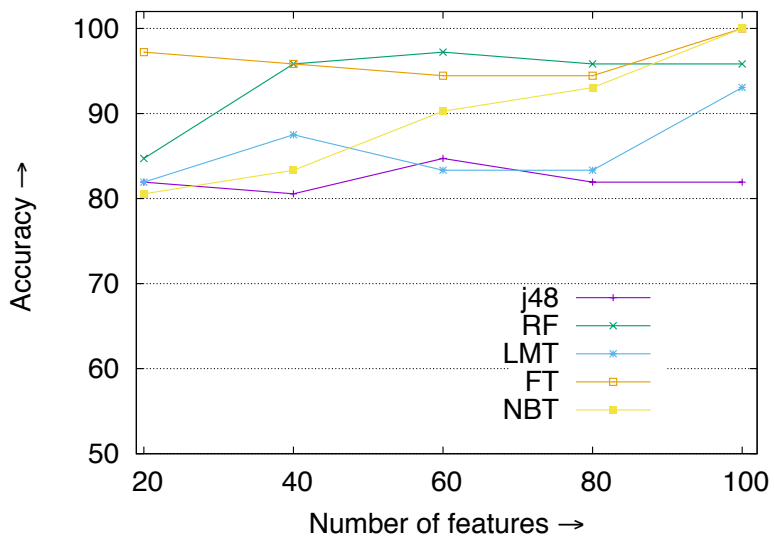

Figure 5: *Detection accuracy obtained by the classifiers on group 1.*

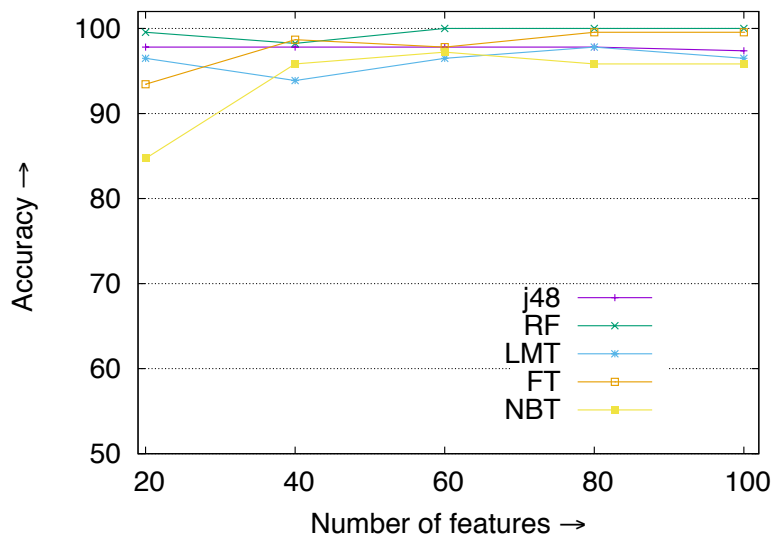

Figure 6: *Detection accuracy obtained by the classifiers on group 2.*

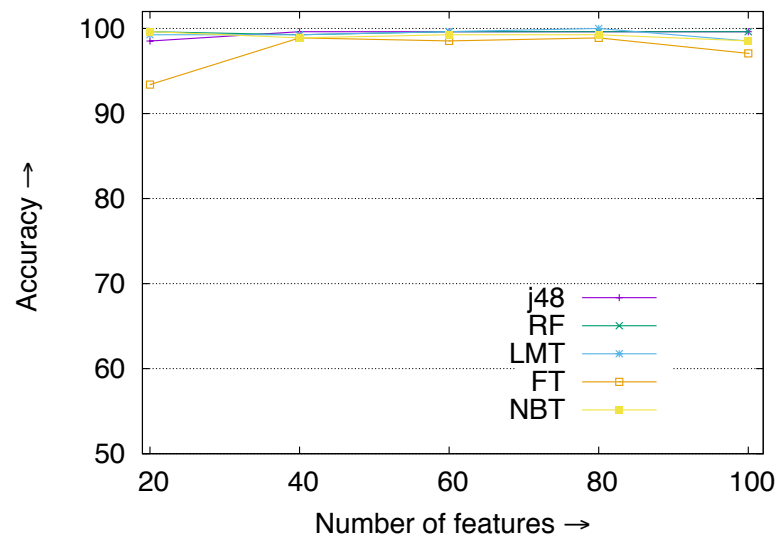

Figure 7: *Detection accuracy obtained by the classifiers on group 3.*

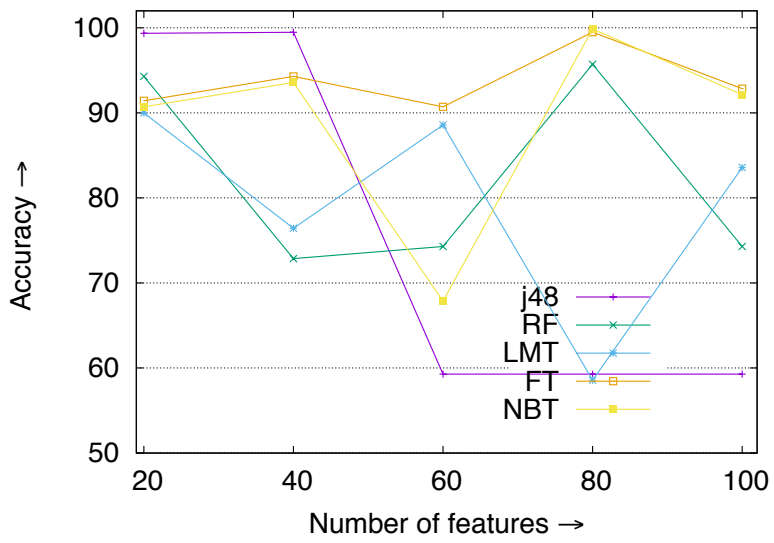

Figure 8: *Detection accuracy obtained by the classifiers on group 4.*

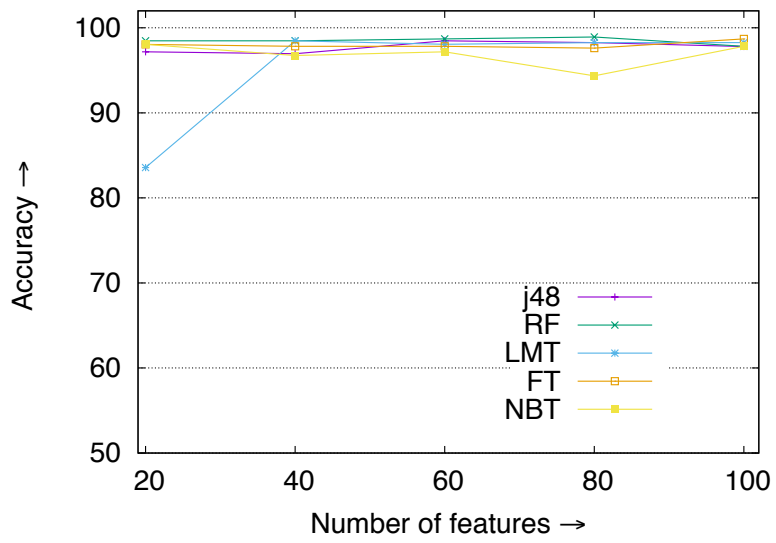

Figure 9: *Detection accuracy obtained by the classifiers on group 5.*

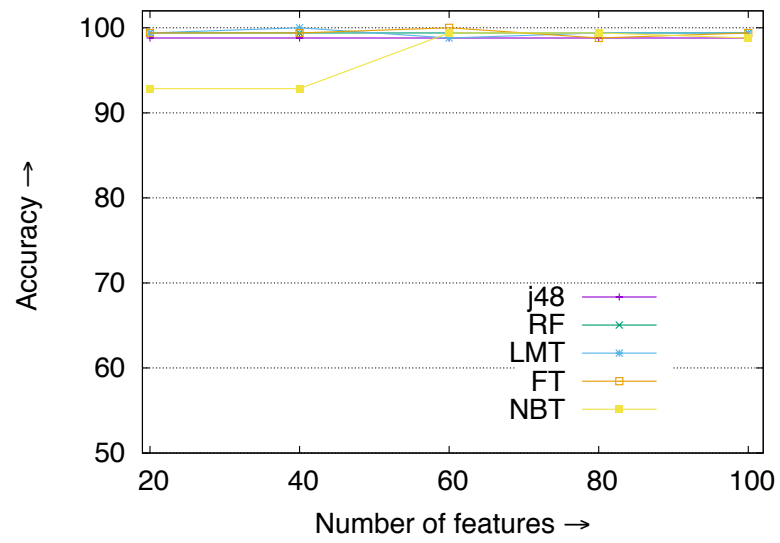

Figure 10: *Detection accuracy obtained by the classifiers on group 6.*

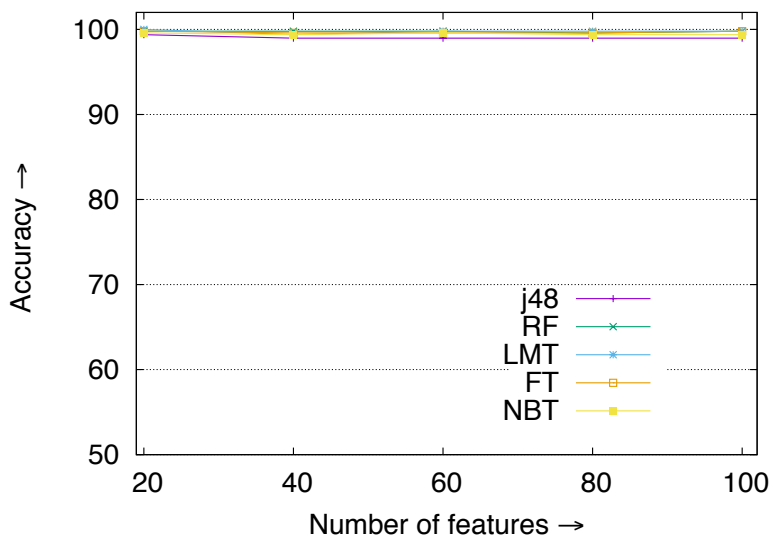

Figure 11: *Detection accuracy obtained by the classifiers on group 7.*

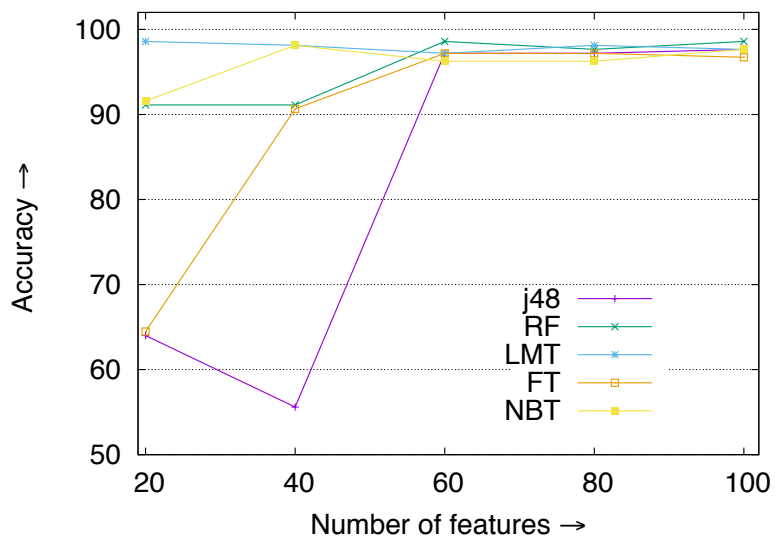

Figure 12: *Detection accuracy obtained by the classifiers on group 8.*

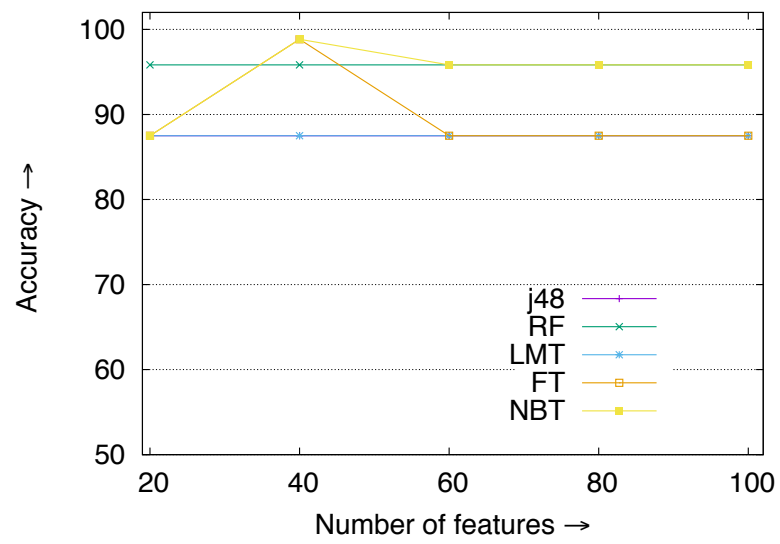

Figure 13: *Detection accuracy obtained by the classifiers on group 9.*

To train the classifiers, first we have used the feature selection algorithm (1) to find the promising features for the detection of malwares for every partitioned group 1-9 and then trained the classifier. To measure the effectiveness of the five classifiers we used different number of features (20, 40, 60, 100) for the classification and monitored the accuracy of classifiers for each groups with providing the respective testing executables of the group. The detection accuracy of the classifier is calculated by the equation.

$$Accuracy(\%) = \frac{TP + TN}{TM + TB} \times 100 \tag{1}$$

where,

TP→ True positive, the number of malwares correctly classified.

TN→ True negative, the number of benign correctly classified.

TM→ Total number of malwares.
TB→ Total number of benign.

The plots (5 - 13) shows the performance of classifiers for each group with respect to different numbers of features provided to the classifiers. We took the best required features from each cluster and find that all the classifier overall performed with correctness of more than 98.21% of detection accuracy. NBT performed better then the other with 99.11% of detection accuracy (shown in fig. 4).

## 5. Conclusion

From the literature, we found that the obfuscation techniques in malware generation are becoming crucial for the classical signature matching approaches used in malwares detection systems. In order to be efficient in analyzing and classifying large amounts of executables for detection of unknown or metamorphic malwares, we have studied the malwares generated by three popular malware generator kits. We found the variation of size of malwares generated by these kits does not vary much. Thus on the basis of findings, we partitioning executables in 9 groups using Optimal k-Means Clustering algorithm on the sizes of malwares as a criteria. Then we discussed a statical analysis approach to detect unknown malwares for windows operating system by selecting features for each group separately by feature selection approach based on the difference of opcode occurrence among benign and malwares executable. We used a well known classifiers viz. Random forest, J48, LMT, FT and NBT for the classification of malwares and found that all of them perform with more than 98% of detection where as NBT give highest accuracy of 99.11%. The experiments suggest that the research direction is promising and the possible future work can be in-depth study of malware generator kits by which we can able to keep a family of malware in one partitioned group and the classifier can attain the detection of unknown malwares with more accuracy.

## Acknowledgements

Mr. Ashu Sharma is thankful to BITS, Pilani, K.K. Birla Goa Campus for the support to carry out his work through Ph.D scholarship No. Ph603226/Jul. 2012/01. We are also thankful to IUCAA, Pune for providing hospitality and computation facility where part of the work was carried out.